\documentclass[useAMS,usenatbib]{mn2e}

\newcommand{\msun}{M$_{\sun}$}
\newcommand{\kms}{km s$^{-1}$}
\newcommand{\msuns}{M$_{\sun}~$}

\newcommand{\mnras}{MNRAS}
\newcommand{\apj}{ApJ}
\newcommand{\apjl}{ApJ}
\newcommand{\aj}{AJ}
\newcommand{\aap}{A\&A}
\newcommand{\araa}{ARA\&A}

\newcommand{\nbody}{{\it N}-body~}

\usepackage{graphicx,amssymb,amsmath}

\title[The rapid dispersal of low-mass virialised clusters]
{The rapid dispersal of low-mass virialised clusters}
\author[Moeckel et al.]{Nickolas Moeckel$^{1}$\thanks{E-mail:nickolas1@gmail.com}, Christopher Holland$^{2}$, Cathie J. Clarke$^{1}$, \& Ian A. Bonnell$^{3}$\\
$^{1}$Institute of Astronomy, University of Cambridge, Madingley Road, Cambridge CB3 0HA\\
$^{2}$Fitzwilliam College, Storey's Way, Cambridge CB3 0DG\\
$^{3}$SUPA, School of Physics and Astronomy, University of St Andrews, North Haugh, St Andrews, Fife, KY16 9SS}

\begin{document}

\date{Accepted XXX. Received XXX; in original form XXX}

\pagerange{\pageref{firstpage}--\pageref{lastpage}} \pubyear{XXXX} 

\maketitle

\label{firstpage}

\begin{abstract}

Infant mortality brought about by the expulsion of a star cluster's natal gas is widely invoked to explain cluster statistics at different ages. While a well studied problem, most recent studies of gas expulsion's effect on a cluster have focused on massive clusters, with stellar counts of order $10^4$. Here we argue that the evolutionary timescales associated with the compact low-mass clusters typical of the median cluster in the Solar neighborhood are short enough that significant dynamical evolution can take place over the ages usually associated with gas expulsion. To test this we perform \nbody simulations of the dynamics of a very young star forming region, with initial conditions drawn from a large-scale hydrodynamic simulation of gravitational collapse and fragmentation. The subclusters we analyse, with populations of a few hundred stars, have high local star formation efficiencies and are roughly virialised even after the gas is removed. Over 10 Myr they expand to a similar degree as would be expected from gas expulsion if they were initially gas-rich, but the expansion is purely due to the internal stellar dynamics of the young clusters. The expansion is such that the stellar densities at 2 Myr match those of YSOs in the Solar neighborhood. We argue that at the low-mass end of the cluster mass spectrum, a deficit of clusters at 10s of Myr does not necessarily imply gas expulsion as a disruption mechanism.

\end{abstract}

\begin{keywords}
open clusters and associations: general -- stars: formation -- stellar dynamics
\end{keywords}

\section{Introduction}
\label{introduction}
A common expression of the star formation paradigm is some variation of ``most stars form in clusters", often followed by a citation to the seminal review of \citet{lada03}. Recent thorough {\it Spitzer} surveys of young stellar objects in the 0.5 kpc radius volume around the Sun \citep{bressert10} reveals a smooth distribution of surface densities, and suggests that a clustered star forming region is probably the high density tail of a log-normal distribution of star forming densities \citep{elmegreen01}. A specific definition or density threshold of a `cluster'\footnote{That is to say, a definition beyond ``I know it when I see it" \citep{stewart64}; for example, \citet{bressert10} choose 200 YSOs pc$^{-2}$ and find 26 per cent of local stars are born above this density, while \citet{lada03} use a definition equivalent to about 3 YSOs pc$^{-2}$ and find that about 90 per cent of stars meet that criteria.} is thus needed to quantify the number of stars born in a clustered environment. 

Regardless of the total fraction of stars born in embedded clusters, gas-free star clusters exist and are observable at large distances, and their statistics must be explained by star formation theories. A potentially large influence on these statistics is the effect of gas removal on an embedded cluster. If star formation proceeds at some rate (that need not be constant in time), the final star formation efficiency (SFE, here defined as the fraction of gas mass converted into stars) of a collapsing and fragmenting clump in a molecular cloud will be determined by the time integrated rate up to the point when gas is expelled, probably via the radiation and winds occurring over the life and death of the more massive stars in the cluster. A typical efficiency may be 10s of per cent of the gas turned into stars; the removal of a potentially large fraction of the mass can, depending on the spatial relationship between the gas and stars and the gas expulsion timescale, unbind a previously bound system.

\begin{figure*}
 \includegraphics[width=180mm]{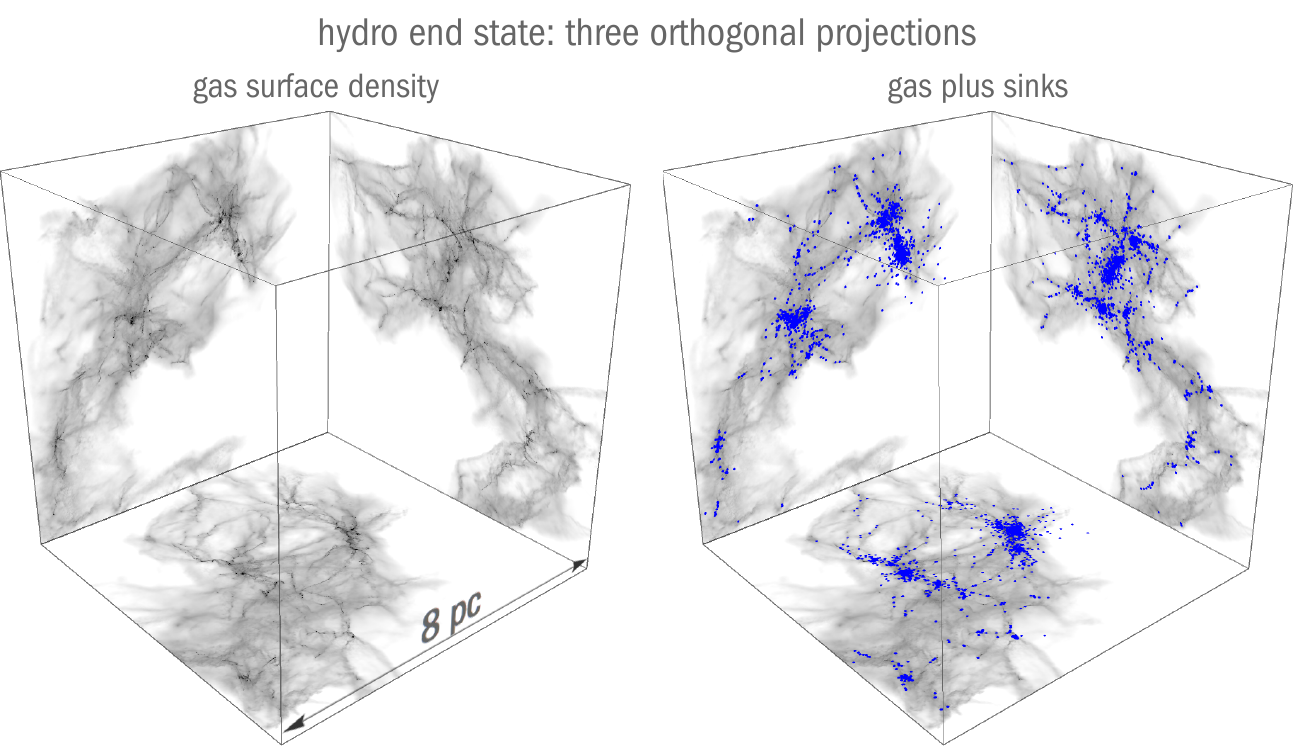}
 \caption{The end state of the hydrodynamic simulation presented in \citet{bonnell08}. At left we show three orthogonal projections of the logarithm of the surface density. At right we include the positions of the sink particles as blue points. These sink positions provide the initial conditions for our \nbody integrations.}
  \label{HydroEndState}
\end{figure*}

The effect of gas removal has been studied in great detail over the past several decades. \citet{hills80} laid out the fundamental analytical theory for systems with gas and stars sharing the same spatial distribution, suggesting that a SFE of greater than 50 per cent is required to end up with a bound system. Numerical and semi-analytic modeling showed that with slow gas expulsion a bound cluster can remain with efficiencies of around 20 per cent \citep{tutukov78,elmegreen83,mathieu83,verschueren89,geyer01,baumgardt07}, and that with rapid expulsion a bound core may remain even as most of the stars are released \citep{lada84,goodwin97,adams00,geyer01,kroupa01,boily03a,boily03,goodwin06,baumgardt07}. While most of this work parameterises the effect of gas expulsion using the SFE, the concept of an effective SFE, \citep[e.g.][]{verschueren89,goodwin06}, allows for some consideration of locally high star formation efficiency even as the global efficiency remains low, an important point to bear in mind when using more natural initial conditions \citep{moeckel10} or including processes that may produce differing distributions for the stars and gas \citep[e.g.][]{moeckel11,smith11}.

The prime motivation for the great deal of attention paid to gas expulsion is its presumably important role in determining whether a star forming region will become a bound cluster or an unbound association. The sample (and cluster definition) of \citet{lada03} implies that by 10 Myr less than 10 per cent of formerly embedded clusters with masses in excess of 150 \msuns are still in existence; cluster disruption due to gas expulsion is a tempting explanation for this apparent infant mortality. While most recent numerical studies of gas expulsion have focused on massive young clusters with $10^4$ stars or more, \citet{adams06} point out that the median cluster population in the \citet{lada03} sample is about 300 stars. At these relatively low numbers, applying the results of studies tailored to a much more populous cluster may suppress important effects due to the $N$ dependence of the timescales in stellar dynamics.

The goal of this paper is to make the point that collisional stellar dynamics\footnote{Note that by `collisional' we are not referring to physical collisions, but rather the regime of stellar dynamics in which interactions between pairs of stars are important. Throughout the paper, we are discussing collisional dynamics.} can play an important role in the early expansion and dispersal of star clusters for clusters of modest mass, of order $10^2$ \msun. In section \ref{timescalesection} we lay out the physical argument and why the effect is largely overlooked, and in section \ref{numericalsection} we explicitly demonstrate the importance of stellar dynamics via numerical simulation.

\section{Timescales and previous work}
\label{timescalesection}

\begin{figure*}
 \includegraphics[width=170mm]{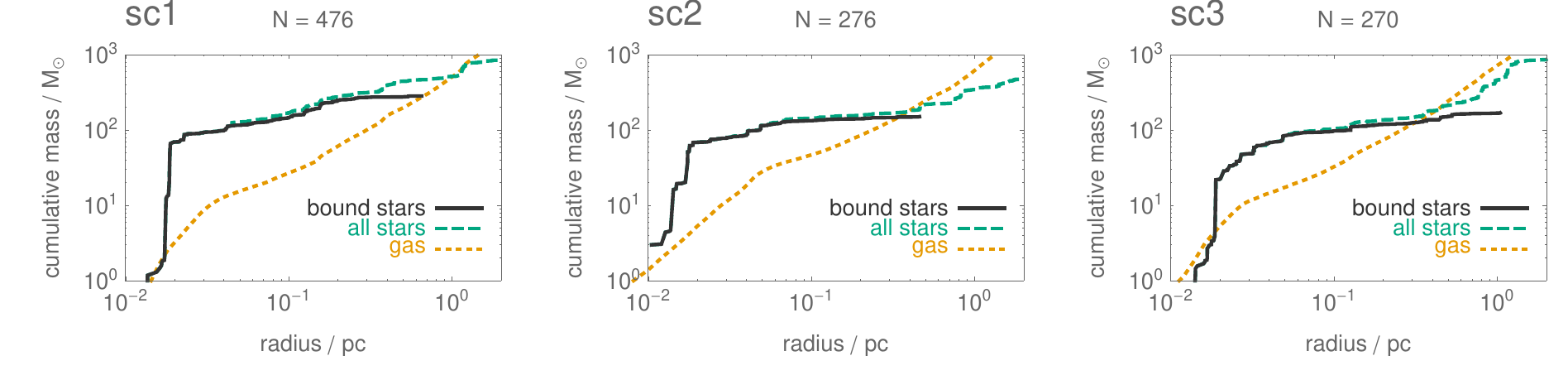}
 \caption{The cumulative mass of stars and gas within 2 pc of our subclusters at the end of the hydro simulation, with radii centered on the clusters. We show both bound stars (i.e. those identified as subcluster members) and all stars in the simulation. These subclusters are dominated by stellar mass.}
  \label{GasPoorClusters}
\end{figure*}

In general, recent numerical studies of gas expulsion\footnote{See the list of references in Section \ref{introduction}.} from star clusters take the following form: i) set up a cluster model, either embedded in an external potential representing the gas or with initially supervirial velocities; ii) if using an external potential, remove the potential over some timescale; iii) allow the stars that remain bound after this stylized gas expulsion to re-equilibrate, and analyse the results. The timescale for the remnant to re-equilibrate is 10s of initial crossing times \citep[e.g.][]{geyer01,baumgardt07}. The chosen timescale for gas expulsion typically ranges from instantaneous to 10s of crossing times. Implicit in this plan is the requirement that no other mechanism besides gas expulsion is altering the cluster structure over the timescales being investigated.

Assuming virial equilibrium, the timescale for stellar dynamics to alter the cluster structure is some multiple of the relaxation time, defined following \citet{spitzer87} as
\begin{equation}
\label{trh}
t_{rh} = 0.138 \frac{N^{1/2} r_{hm}^{3/2}}{\left ( G m \right ) ^{1/2} {\rm ln} \left ( \gamma N \right) },
\end{equation}
while the crossing time of the cluster is 
\begin{equation}
\label{tcross}
t_{cr} = \frac{ r_{hm}^{3/2}}{\left ( G m N\right ) ^{1/2}}.
\end{equation}
Here $N$ is the number of stars in the cluster, $r_{hm}$ is the half-mass radius, $m$ is the mean stellar mass, and the $\gamma$ is usually taken in the range $\gamma = 0.02$--$0.11$; the smaller value is appropriate for a full mass spectrum, and the larger for a single massed system \citep{giersz96}. For small-$N$ systems the value of $t_{rh}$ depends quite strongly on the chosen $\gamma$.

Most recent studies of gas expulsion have made initial condition and computational choices that drive the relaxation time to large values, with the effect of making detailed collisional stellar dynamics largely ignorable. The first choice is to focus on massive clusters, with $N$ of order $10^4$ stars or more. This makes the relaxation time long compared to the age of a cluster when gas expulsion likely occurs, typically assumed to be less than a few million years. For example: a cluster with $r_{hm} = 0.5$ pc, a mean stellar mass of 0.4 \msuns, and $2\times10^5$ stars has $t_{rh} \sim 30$ Myr, while $t_{cr}\sim0.2$ Myr. Even after the 10s of initial crossing times required to regain equilibrium after gas expulsion, the cluster is still only a few Myr old, and much less than an initial relaxation time old. This separation of timescales is why stellar dynamics have generally been unimportant in previous work on gas expulsion\footnote{In an age of more limited computational resources, the large-$N$ choice was not possible, and it is interesting to note that \citet{lada84}, using unsoftened stellar potentials, begin to see mass segregation in their simulations of 50--100 stars with a mass function.}, although as we discuss below even an order of magnitude separation between the relaxation time and the time to re-equilibrate may not suffice to isolate the cluster from important collisional effects. 

Two other choices contribute to lengthening the relaxation time. The first, purely numerical, is the use of smoothed stellar potentials. These artificially lengthen the relaxation timescale by reducing the effects of two-body relaxation. The standard tools of modern stellar dynamics tend not to use softening, so this is not as common as it once was. The second is more frequently seen in recent studies, and is the use of equal mass stars. The effect of this choice is not immediately obvious by glancing at the definitions of the crossing and relaxation times, which depend only on the mean stellar mass; the importance of a mass function lies in the scaling of the coefficients in front of those timescales.

The time for a cluster to reach core collapse, when energy generated by hard binaries' interactions with other cluster members begins to heat the cluster and drive expansion, can vary by orders of magnitude depending on the mass of the most massive stars. For a standard Plummer sphere model, core collapse in an equal mass system occurs at about $15 t_{rh}$. When a broad mass spectrum is used, core collapse can occur at times less than $0.1 t_{rh}$ \citep{gurkan04}. This is a result of mass segregation, which takes place on a timescale inversely proportional to the mass of the object being segregated. If massive stars quickly find their way to the center they can effectively accelerate core collapse as they transfer energy outwards and sink to the cluster center in a futile effort to regain equilibrium with their lower-mass cluster siblings \citep{spitzer69}.

Consider now a cluster with $N=200$ and $r_{hm}$ = 0.1 pc. The crossing time is about 0.17 Myr, and the relaxation time about 1.2 Myr. If there is enough of a mass spectrum present that core collapse will occur after a relaxation time or less, considerable dynamical evolution will take place on the scale of a few crossing times. The separation of timescales disappears, and gas expulsion and stellar dynamical effects may occur simultaneously. While a cluster of 200 stars may seem unimportant, recall that this number is similar to the median embedded cluster in the solar neighborhood by the definition of \citet{lada03}, and furthermore in a hierarchical merging cluster formation scenario \citep[e.g.][]{bonnell03}, many $N=200$ star subclusters may merge to form something on the scale of larger local star forming regions like the Orion Nebula Cluster.

\section{Demonstration via numerical simulation}
\label{numericalsection}
In order to demonstrate the potential importance of stellar dynamics in modestly populous young clusters, we apply the approach of \citet{moeckel10} to the cluster presented in \citet{bonnell08}. In this type of experiment the end state of an {\it ab initio} hydrodynamic star formation simulation is used to generate initial conditions for a gravitational $N$-body simulation. 

\begin{figure}
 \includegraphics[width=84mm]{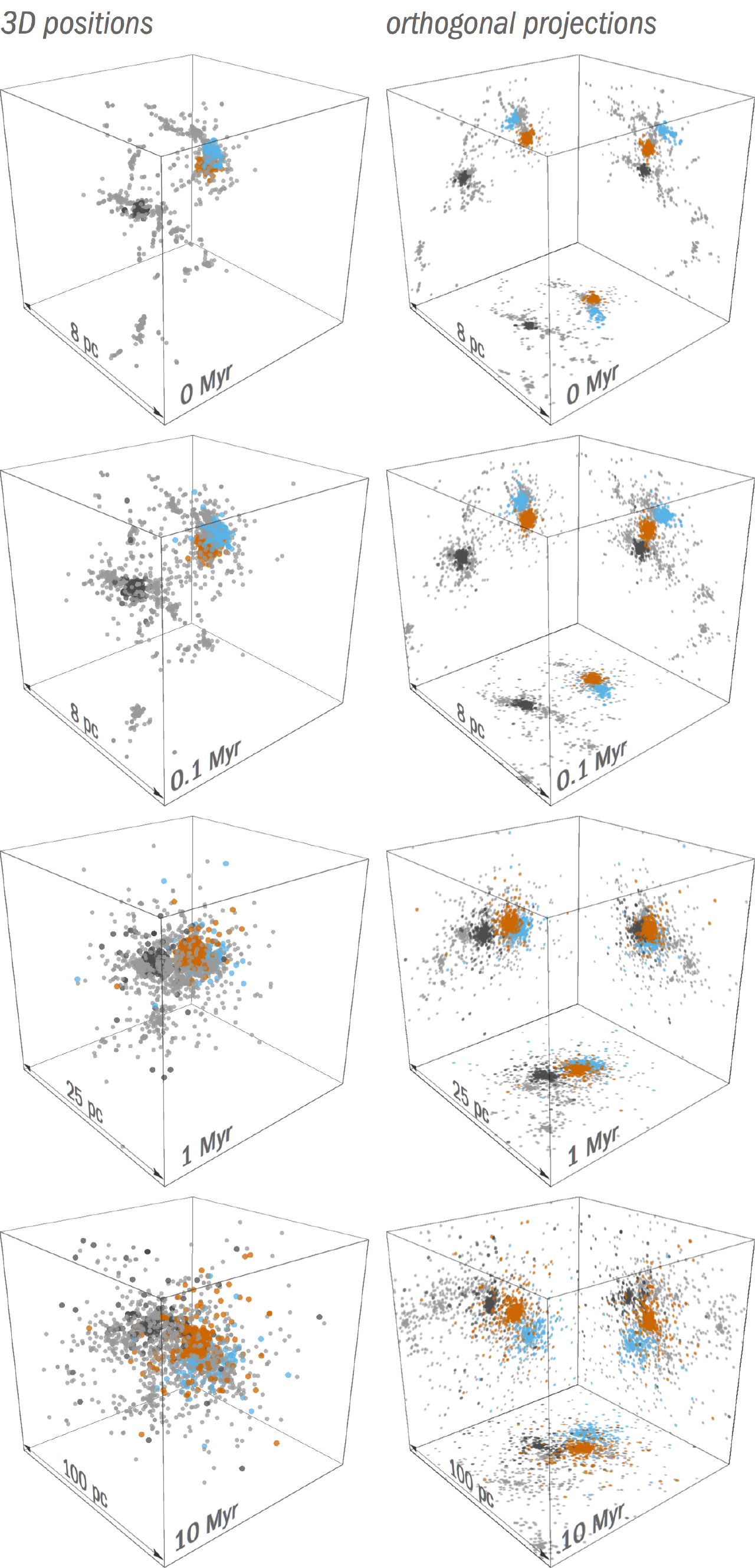}
 \caption{The evolution of the subclusters after the gas has been instantaneously removed. Stars that are initially bound members of the three most massive subclusters are coloured orange (sc1), black (sc2), and blue (sc3). The rest of the stars are gray. The left column shows the stars in 3 dimensions, and the right column shows their orthogonal projections onto the cube walls. Note the changing length scale.}
  \label{FullClusterGrid}
\end{figure}

\subsection{Initial conditions and computational concerns}
Before discussing the evolution of the star forming region, we briefly describe the numerical techniques used to generate the data for this paper.

\begin{figure*}
 \includegraphics[width=170mm]{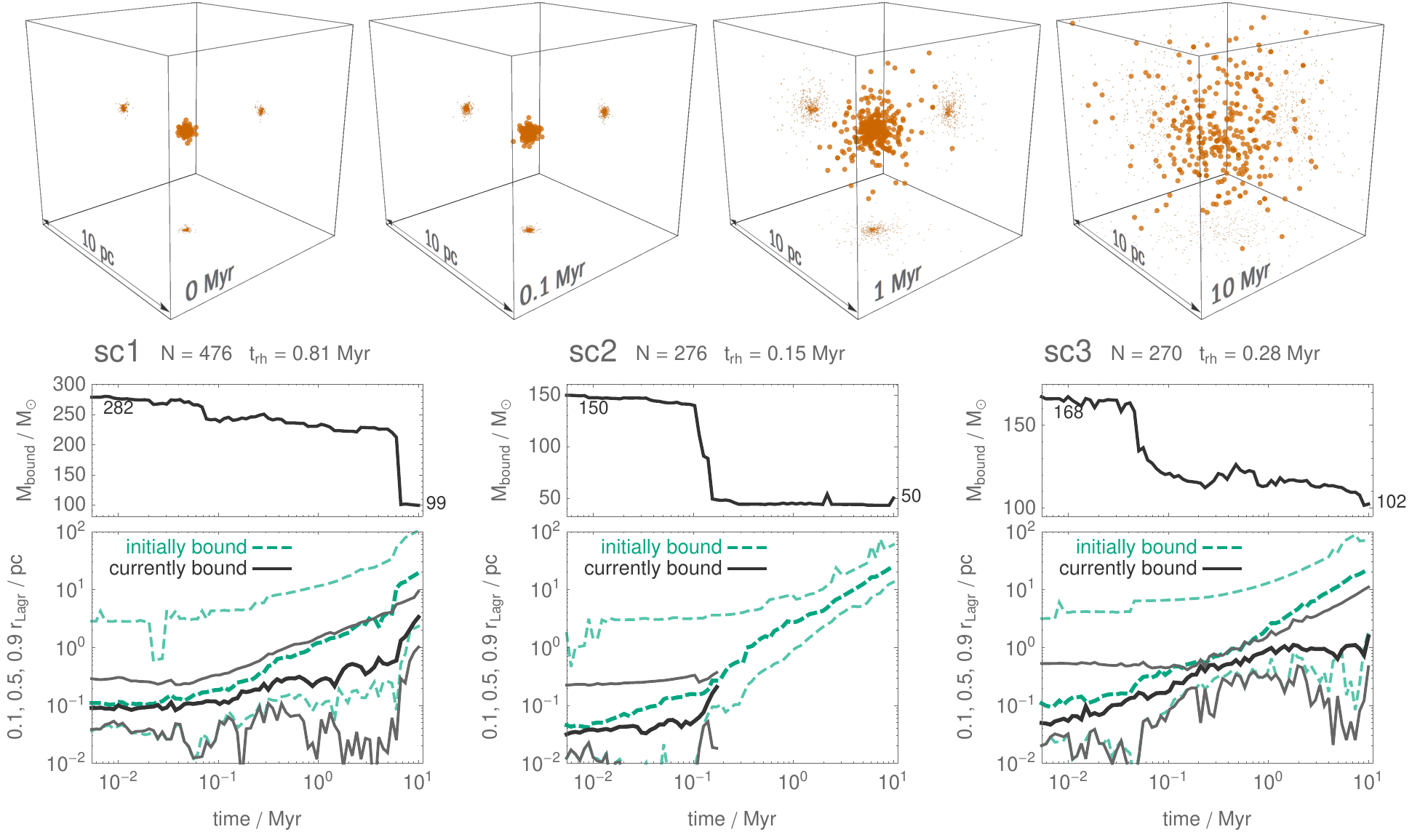}
 \caption{The top row of 10 pc boxes show the positions of all the stars initially bound to sc1, centered on the subscluster center of mass. The projections onto the walls of the box are also shown. The other two subclusters evolve visually similarly. The bottom row of plots shows the structural evolution of each subcluster. In their top panels the plots show the mass of bound stars as a function of time. The numbers at the ends of the curves are the initial and final bound mass. The lower panels show the 0.1, 0.5, and 0.9 Lagrangian radii of all the stars that were initially bound (blue-green dashed lines) as well as only those stars still bound at a given time (gray lines). The 0.5 Lagrangian radius is the thickest line in each set. `Currently bound' lines are only plotted when more than 20 stars remain bound.}
  \label{LagrRadPlot}
\end{figure*}

\subsubsection{Initial conditions from the hydrodynamic simulation}
While \citet{bonnell08} and \citet{bonnell11} should be referred to for full details of the star forming hydrodynamic calculation, the model and numerical approach are as follows: a $10^4$ \msuns cylindrical molecular cloud, initially seeded with a turbulent velocity field, was simulated using smoothed particle hydrodynamics. The calculation utilised 15.5 million particles at two levels of mass resolution, a barotropic equation of state modeled after \citet{larson05}, and sink particles \citep{bate95} to model star formation. The sink particles are a numerical construct used to remove increasingly dense regions of collapsing gas from a calculation, under the assumption that gas flowing into the sink region will be accreted onto a star. The compromise is that physics taking place inside the sink radius is lost. In the Bonnell et al. simulation the sink radius is 40 au, and this is also the distance at which gravitational interactions between sinks are softened. A study of the hierarchical merger-driven growth of stellar structures and an analysis of the dynamical state of this simulation are provided in \citet{maschberger10} and \citet{kruijssen12}, respectively.

After about $6.5 \times 10^5$ yr, the end state of the hydrodynamic calculation is a collection of 2542 sink particles, with masses from 0.12 -- 30.5 \msun. The total mass of the stars is 1586 \msun, thus a global star formation efficiency of about 16 per cent. However, the distribution of gas and stars is very structured, and local SFEs are much higher. In Figure \ref{HydroEndState} we show projections of the gas and the sinks at this time. The positions of the sinks provide our initial conditions for \nbody simulations, and from this point onwards we will identify them as stars. They have physical radii corresponding to their mass, their gravitational interactions are unsmoothed, and they undergo stellar evolution assuming Solar metalicity \citep{hurley00}. The gas is removed from our consideration; this instantaneous 'gas expulsion' is the most extreme possible parameterisation of gas loss and should result in a clear signal of any gas expulsion driven cluster expansion.

\subsubsection{Longer term evolution with N-body simulations}
In order to evolve the stellar system forward in time, we use the \nbody code {\sc nbody6} \citep{aarseth00,aarseth03}, which uses a 4th order Hermite scheme coupled to KS regularisation \citep{kustaanheimo65} to handle binary systems. For simplicity we evolve the system in isolation. As mentioned above, stellar evolution and collisions are enabled. The integration proceeds for 10 Myr from the end of the hydrodynamic simulation. All times quoted take the beginning of the \nbody integration as $t=0$.

Because we are dealing with a modest number of stars, it would be preferable to have multiple sets of statistically identical initial conditions to work with in order to disentangle the chaos of small-$N$ dynamics from general trends in a system's evolution. We are restricted to the single hydrodynamic run, however. To allay this concern we run 8 additional mild randomisations of the initial conditions. These are constructed by jiggling the velocity of each star by a small amount. Specifically, a star with velocity ${\bf v}$ has a randomly oriented vector of magnitude $0.01 | {\bf v} |$ added to it. This is sufficient to scramble the precise trajectory of a star through a cluster without destroying the bulk velocity structure of the initial conditions. The results of these randomisations will be discussed when below the discussion of the main simulation.

\subsubsection{Identification of a few interesting subsclusters}
We focus on the evolution of all the stars in the star forming region, as well as the three largest subclusters. These were identified by using an agglomerative hierarchy finding routine, which builds a cluster hierarchy in which objects are merged using Ward's criterion\footnote{At each step of the hierarchy construction, the pair of objects being merged is the pair that minimizes the increase of the variance in positions of the new object's members.}. Starting from the individual stars, the merging tree continued until there were six groups of stars. This procedure serves to divide the region up into groupings as might be done by human eye. The three most populous of these were identified as subclusters of interest.

The membership of the clusters was then restricted to bound members only, following the method used in \citet{baumgardt02}. To do this the potential energy of each star in the subcluster was calculated relative to all the other potential subcluster members, as well as the kinetic energy in the centre of momentum frame of the subcluster. Those stars with negative total energies are provisionally bound. The process is then iterated, with the potential energy of each star calculated only with respect to the bound members, until a stable bound population is reached. The first step of the cluster finding is thus just a rough guide to restrict the boundedness calculation to a reasonable region.

Each of these subclusters then consists of stars that are bound to themselves {\em without any contribution from the gas}. While it may seem like cheating to ignore stars that may become unbound by the gas expulsion, it is precisely the behaviour of the remaining bound subclusters that we wish to study. In any case, as discussed by \citet{kruijssen12}, the subclusters in this simulation are approximately virialized without taking the gas into consideration, implying that they are gas poor (i.e. that the local star formation efficiency is high). We note that similar high stellar density, gas-poor clusters are also found in simulations of star formation using an AMR code \citep{girichidis11,girichidis12}. We show this explicitly in Figure \ref{GasPoorClusters}, plotting the cumulative mass of gas and stars within 2 pc of the center of each of our bound subclusters. The stars dominate the mass in these regions, and restricting ourselves to bound members actually doesn't exclude very many stars except at the outer edge of the cluster where the gas mass begins to become significant. We identify the three subclusters as: sc1, with 476 stars; sc2, with 276 stars; and sc3, with 270 stars. These three bound subclusters thus contain about 40 per cent of the stars in the simulation.

\subsection{Evolution of the region after gas removal}

In Figure \ref{FullClusterGrid} we show the positions of the stars at four times. The members of sc1 are in orange, sc2 in black, and sc3 in blue; the remaining stars are shown in gray. If pressed to summarise the region's global evolution with a single word, `expansion' would be appropriate. The cluster goes from being comfortably enclosed by a 8 pc box to a 100 pc box over the 10 Myr integrations. At first glance this might seem to be a simple matter of expansion due to the removal of most of the mass in the system. Certainly some of the stars are unbound from the region's centre of mass and sent streaming away due to the gas expulsion. However, as shown in Figure \ref{GasPoorClusters} and as discussed by \citet{kruijssen12}, the subclusters we focus on are gas poor and virialized. Gas expulsion is not expected to affect them very much, and recall that just the three we analyse here contain nearly half the stars in the simulation. In the right hand panels of Figure \ref{FullClusterGrid}, showing the projections of the stars onto the cube walls, note that the three subclusters remain identifiable as separate entities throughout the 10 Myr, though they do expand from their original size. The overall evolution of this region is due to two separate physical effects: first, the dispersal of the subclusters relative to each other, which is caused by gas expulsion; on top of that, the expansion of each of the subclusters due to stellar dynamics, {\em not} gas expulsion. The combination of virialised subclusters in a cluster that is not necessarily virial is in contrast to most previous studies of structured initial conditions, such as \citet{goodwin04b}, where the virial state of the system is parameterised and is constant across all scales.

In Figure \ref{LagrRadPlot} we show the positions of the stars of sc1 in time as it expands to fill a 10 pc box, as well as plots showing the bound mass and Lagrangian radii of each subcluster. Lagrangian radii enclose a fixed mass fraction of a cluster, and their behaviour over time are diagnostic of the physical processes driving a cluster's evolution. We discuss the features of these plots individually.

\begin{figure}
 \includegraphics[width=84mm]{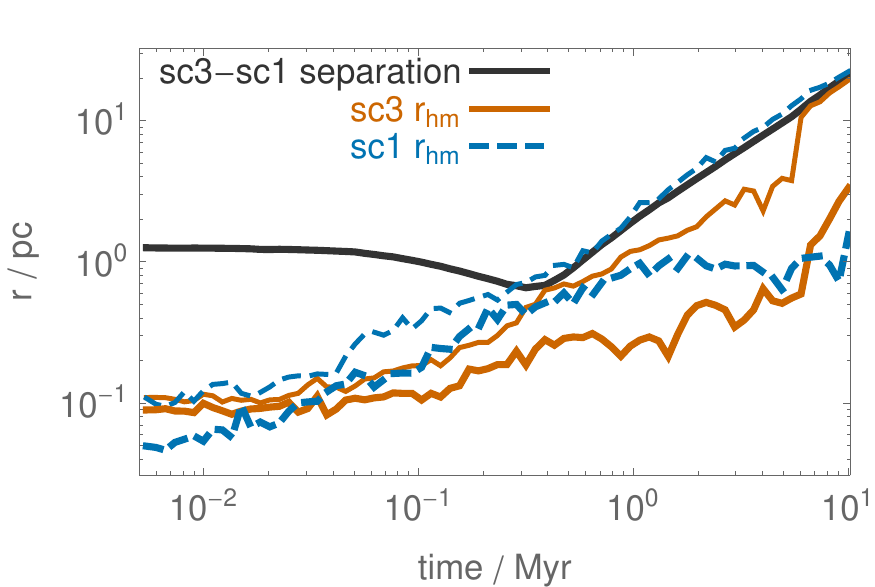}
 \caption{The separation between sc3 and sc1. These subclusters undergo a hyperbolic encounter at about 0.3 Myr. The distance between their density centres are shown as the black line. The half-mass radii of their bound stars (thicker lines) and all initially bound stars (thinner lines) are plotted as well.}
  \label{sc31sep}
\end{figure}

\subsubsection{Expansion of sc1}
The most massive subcluster sc1 undergoes the most orderly expansion, since by virtue of its larger population it is less susceptible to small-$N$ chaos yielding large changes to the cluster structure. For about the first $10^5$ yr very little happens to the subcluster. Some stars are lost and some expansion takes place, but it is only after this time that general expansion of all the Lagrangian radii occurs. With a half-mass relaxation time of 0.86 Myr, the time at which real expansion begins is comparable to 0.1 $t_{rh}$. This is probably somewhat coincidental though, since this subcluster contains the most massive star in the simulation, which via dynamics and the star formation process is already part of a massive binary near the subcluster center when the \nbody integrations begin. From the point of view of the \nbody integration, this subcluster is not just primordially mass segregated, but is in primordial core collapse. 

The slowly accelerating expansion of the cluster's bound membership (i.e. the Lagrangian radii begin flat and steepen to an approximately steady slope) is characteristic of the onset of self-similar expansion due to an energy source at the cluster centre \citep{henon65}. While the energy in sc2 and sc3 comes from heating by hard binaries in the cluster core, sc1 contains four massive stars with masses greater than about 10 \msun that provide a small contribution from mass loss during their evolution. The large drop in bound mass at about 6 Myr is a result of a supernova involving the 30.5 \msuns star. This event is the dominant shaper of the cluster structure from that point onwards; the loss of most of this star's mass, as well as its $\sim 15$ \msuns companion, frees a large fraction of the remaining stars. In contrast to what we see here, the signal of expansion due to rapid gas expulsion is immediate and decelerating expansion, with Lagrangian radii that are steeper at first and flatten to a steady slope \citep[e.g. figure 6 of][]{baumgardt07}.

The initial 3D velocity dispersion of sc1 (as well as sc2 and sc3) is $\sigma \sim 3.5$ \kms. Assuming for simplicity linear motion, if the stars {\em had} been released at that velocity because of gas expulsion they might be expected to travel a distance $d = \sigma t$; with $t = 10$ Myr, $d \sim 35$ pc. This value is only slightly larger than the final half mass radius of the all the stars initially bound to the subclusters, shown as the dashed blue-green lines in figure \ref{LagrRadPlot}. This is not surprising. The escape speed from the subcluster and its velocity dispersion are both, to within factors of order unity, approximately $(GM/r_{hm})^{1/2}$, with $M$ the subcluster mass. Stars that are unbound by stellar dynamical effects must be traveling at somewhat greater than the escape speed\footnote{The argument holds whether the escape velocity is attained by two-body effects or direct ejection from a binary interaction.}. Since over half the subcluster mass escapes there is a large number of stars streaming away from the sc1's density centre, and the different physical drivers of expansion give similar final sizes. This near equivalency is only possible with low-{\it N} systems like these, where the crossing time and relaxation time are not so different. Then the dynamically escaping stars ($\sim 10$ per cent of the stars per relaxation time) become a significant fraction of the total population on a timescale similar to that on which gas expulsion releases the stars, several crossing times.

\subsubsection{Expansion of sc2}
This subcluster begins its evolution similarly to sc1, but at about $10^5$ yr a small-$N$ interaction involving some of the most massive stars in the cluster core ejects a 10.8 \msuns star with high velocity. The binary responsible for the ejection consists of a 15.2 \msuns and a 14.4 \msuns star. This is not the first time that these three stars had encountered one another, as the 10.8 and 14.4 \msuns stars each take turns in the binary prior to the ejection. The final encounter is energetic enough that the recoiling binary is also ejected, and the loss of 40 \msuns of mass is enough to disrupt the cluster. The cluster begins to freely expand at that point, as seen in the abrupt steepening of all three Lagrangian radii that we plot. The bound mass is formally 50 \msuns comprising 10 stars in 5 stellar systems, but these objects are very weakly bound and do not interact for the remainder of the simulation.

\subsubsection{Expansion of sc3}
While this subcluster likewise experiences a small-$N$ dynamical decay that dislodges a significant number of stars (at about 0.05 Myr), over 100 \msuns of stars remain bound and the evolution proceeds similarly to sc1. This group and sc1 have a relatively close hyperbolic encounter at about 0.3 Myr, with their centres of mass passing just within 0.7 pc of each other. After this encounter their separation increases nearly linearly, but at a rate that nearly matches the expansion of their stellar populations. This is shown in figure \ref{sc31sep}, where we plot the separation of the sc3 and sc1 centres of mass, as well as the half-mass radii of currently and initially bound stars in those subclusters. The match between the separation and the initially-bound half mass radii means that these subclusters are embedded in their mutual escaped detritus; this can be seen in figure \ref{FullClusterGrid}.

\subsubsection{Evolution of the stellar densities}
\begin{figure}
 \includegraphics[width=84mm]{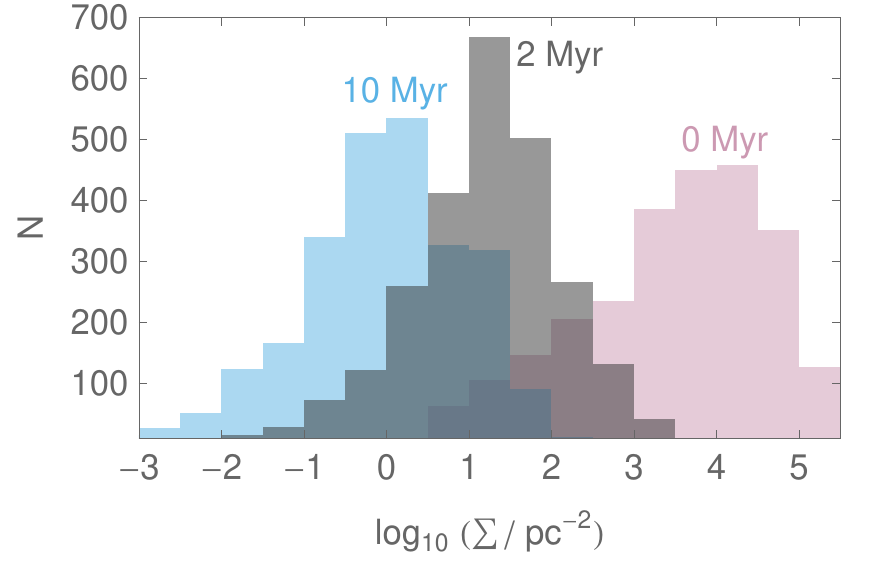}
 \caption{The evolution of the stellar surface density distribution with time. These histograms are using a single projection of the 3D positions to a plane (the projection on the bottom of the cube in Figure \ref{HydroEndState}; using other projections can shift the histograms around by small factors, but the order of magnitude shifts seen are robust.}
  \label{SurfaceDensities}
\end{figure}

\begin{figure}
 \includegraphics[width=84mm]{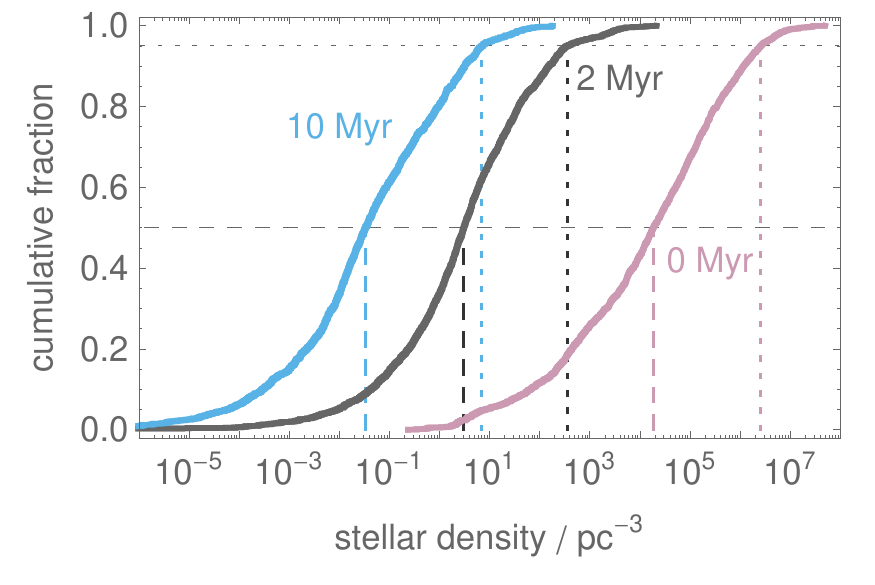}
 \caption{The evolution of the stellar volume density distribution with time. The vertical dashed and dotted lines show the median and 95 per centile of each distribution, respectively.}
  \label{VolumeDensities}
\end{figure}

\begin{figure*}
 \includegraphics[width=180mm]{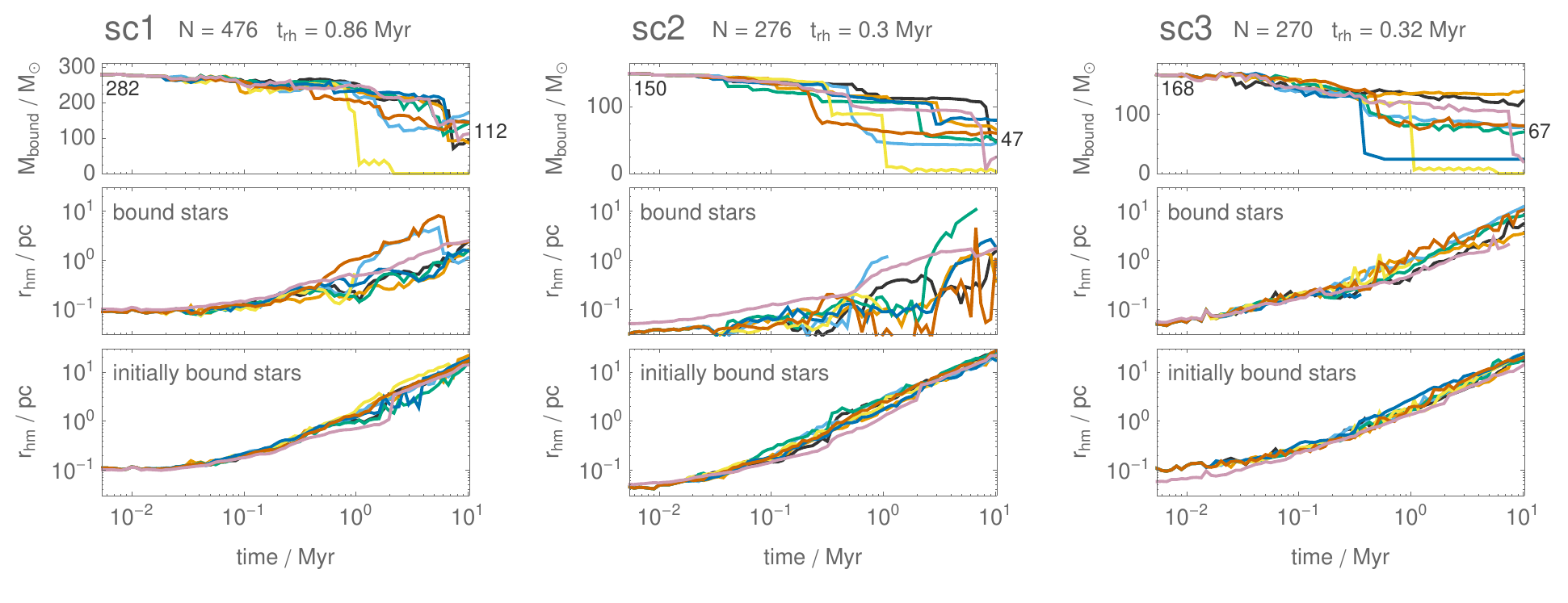}
 \caption{Summary of the results of the randomised initial conditions. We plot the bound mass, the half mass radius of bound stars, and the half mass radius of all originally bound stars for each of the 8 initial conditions. The lines for the bound stars are only shown when at least 20 stars remain bound. The final mass shown is the median of the 8 realisations.}
  \label{RandomComparison}
\end{figure*}

The combined effect of the dispersal of the subclusters relative to each other and the internally-driven expansion of each subcluster is to spread out the stars in space, which is clear in figure \ref{FullClusterGrid}. The initial half mass radius of the entire region is about 1.6 pc, and as \citet{kruijssen12} point out the distribution of YSO surface densities is approximately log-normal with a peak at about 5600 YSOs pc$^{-2}$. While far in excess of the 22 YSOs pc$^{-2}$ that is the peak of the local distribution \citep{bressert10}, the similar shape of the distribution despite a very clustered distribution is noteable. After 2 Myr, the overall half-mass radius of the region is 5.5 pc, and the peak of the stellar surface density distribution is around 20 pc$^{-2}$ (the precise value depends on the projection being analysed). At 10 Myr, the half-mass radius is 24 pc, and the surface density peak has dropped to about 2 pc$^{-2}$. For comparison to figure 1 of \citet{bressert10}, these surface density distributions are plotted in Figure \ref{SurfaceDensities}. The agreement at 2 Myr between very dense initial conditions and the observed distribution is striking. The cumulative distribution of the volume densities at the same times \citep[calculated, like the surface densities, using the method of][]{casertano85} are plotted in figure \ref{VolumeDensities}.

\subsubsection{Variations with randomised initial conditions}
These are small-$N$ subclusters, and chaotic small-$N$ encounters are demonstrably important in shaping the overall evolution of the bound members. We now examine the randomised initial conditions to see how universal the behaviour in the main simulation is. In Figure \ref{RandomComparison} we plot the bound mass and the half mass radius of initially bound and currently bound stars for all 8 randomised cases. Looking at sc1, with two exceptions the half-mass radii of the bound stars are within a factor of about 2 at all times. The outliers are cases where the most massive star is nearly ejected, and because it is so massive the half-mass radius roughly tracks its position; when it is lost as a supernova the half-mass radius returns to the same trend as the other runs. In all the realisations there is some randomness arising from ejections of massive stars, which appears in the bound mass as well as the half-mass radii, but by the time the time the supernova occurs the half-mass radii have reconverged. One random example completely disassociates at around 1 Myr, but most of the sets follow a similar path as our main simulation.

The scatter in bound mass and the half-mass radius of the currently bound stars for sc2 is particularly large. This is the cluster that in our main simulation is dissolved when the most massive stars eject themselves. The large scatter in the bound mass and bound half-mass radius reflect variations of this scenario. While a wide range in the behaviour of the bound quantities are clearly possible with these small-$N$ subclusters, the half-mass radius of all initially bound stars are less affected by the chaotic nature of the systems. Note that the evolution of sc2 in our main simulation is an outlier compared to the randomised set of runs; it drops to a very small bound population earlier than any others. 

\subsubsection{The insignificance of the subclusters' orbits and their interactions}
\begin{figure}
 \includegraphics[width=84mm]{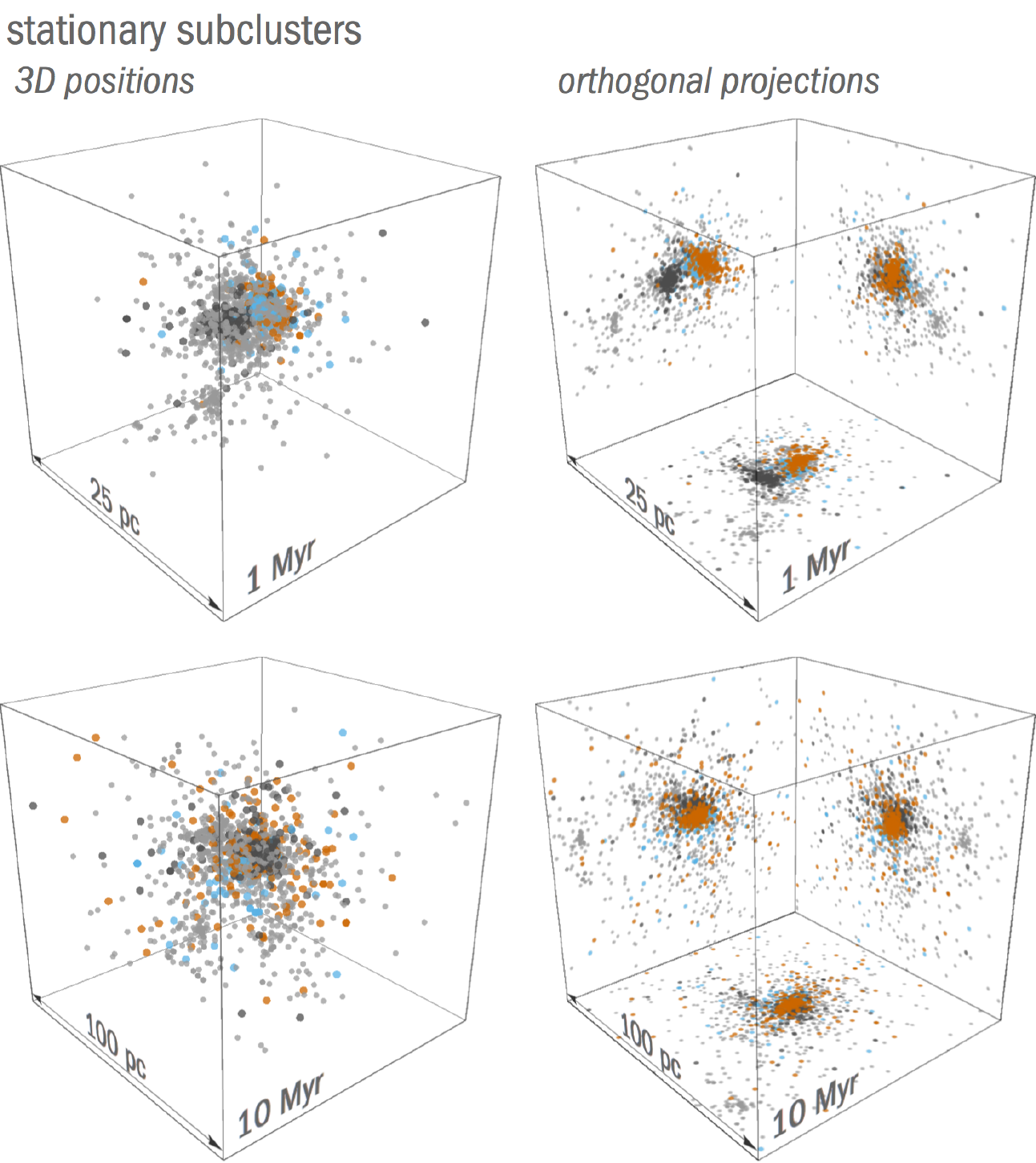}
 \caption{The evolution of the subclusters with the six most populous groups started from rest relative to each other. Stars that are initially bound members of the three most massive subclusters are coloured orange (sc1), black (sc2), and blue (sc3). The rest of the stars are gray. The left column shows the stars in 3 dimensions, and the right column shows their orthogonal projections onto the cube walls. Note the changing length scale.}
  \label{StationaryClusterGrid}
\end{figure}

As we have mentioned, the subclusters are unbound from each other in our initial conditions, a state that is attributable to the removal of the system's gas. How much of the large-scale expansion seen in figure \ref{FullClusterGrid} is due to the increasing separation between the subclusters, and not the stellar dynamical expansion that we claim is dominant? To get a feeling for this we ran an experiment where we have removed the bulk motion of the six most massive subclusters. These encompass the three that we anaylse in detail here, as well as the next three most populous groups. The clusters now start from rest relative to each other; rather than flying apart they are guaranteed to merge at some point.

\begin{figure}
 \includegraphics[width=84mm]{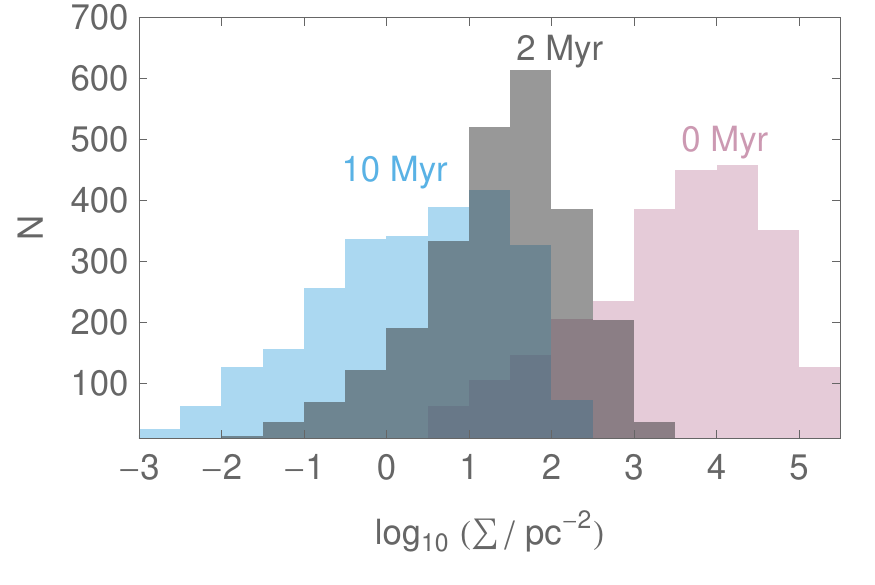}
 \caption{The evolution of the stellar surface density distribution with time, for the run with the subclusters initially stationary.}
  \label{StationarySigmas}
\end{figure}

In Figure \ref{StationaryClusterGrid} we show an abbreviated version of Figure \ref{FullClusterGrid} for this run, showing the stars' positions at 1 and 10 Myr. At 1 Myr, sc1 and sc3 have begun to merge, while sc2 has still not yet joined the center of mass. However, by this point the self-similar expansion of the clusters is already well underway. At 10 Myr the individual subclusters are not distinct, in contrast to our main results; however, the overall scale of cluster expansiion is very similar, as seen by the way that the particles fill their respective boxes at 1 and 10 Myr in Figures \ref{FullClusterGrid} and \ref{StationaryClusterGrid}. This means that the evolution of the surface densities is very similar to the main run, and is shown in figure \ref{StationarySigmas}. At 2 Myr the surface density peaks around 30 rather than 20 YSOs pc$^{-2}$. The overall behaviour of the system is quite similar to the main run despite the very different orbits of the subclusters, lending credence to our claim that the internal dynamics of the subclusters (rather than the cluster to cluster relative motions) dominate the global expansion in the system. The expansion of the subclusters away from each other contributes something like a factor of 2 to the lowering of the surface density, depending on the projection chosen, while stellar dynamical expansion is responsible for the order of magnitude. 

\begin{figure}
 \includegraphics[width=84mm]{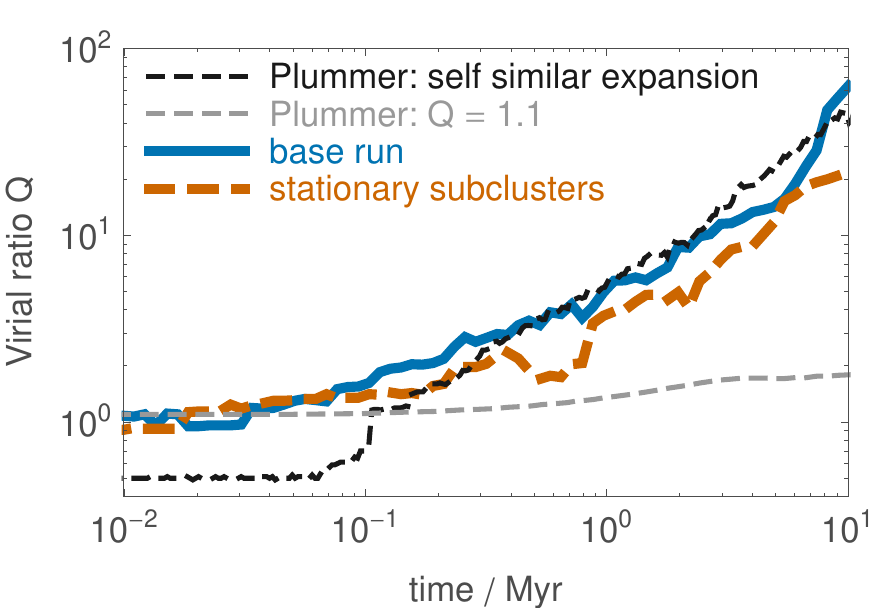}
 \caption{The change to the virial ratio $Q$ with time for our main run (solid blue line) and the run with initially stationary subclusters (dotted orange line). For comparison, the results from Plummer spheres expanding due to different physical mechanisms are also plotted (thinner dashed gray lines).}
  \label{VirialRatios}
\end{figure}

To further illustrate the dominance of binary driven expansion we plot in Figure \ref{VirialRatios} the global virial ratio $Q = E_{kin} / |E_{pot}|$ of the system for the main run and the initially stationary subcluster run. Hard binaries are treated as their center of mass for these calculations, so that their binding energy is not included. The main run begins with $Q = 1.1$, i.e. marginally unbound. The value of $Q$ increases with time as energy from the binaries heats the subclusters. The run with initially stationary subclusters begins with $Q = 0.9$, supervirial but still bound\footnote{The supervirial value of $Q$ is mostly due to the population of stars that is not in virialised subclusters. Because the cluster--cluster velocities and the velocity dispersion within clusters are comparable, halting the subcluster motion does not dramatically lower $Q$.}. The evolution of $Q$ is quite similar. For comparison we plot the result from spherical star clusters evolved in isolation: one that begins virialised at $Q=0.5$ and goes through core collapse into self-similar expansion, and one that begins with $Q = 1.1$ and is not heated by binaries. The virialised run has been scaled to begin expansion at a similar time as our main simulation. The similarity between our runs and the stellar dynamically expanding cluster is clear, as is the dissimilarity to the run with expansion due to only to comparably supervirial initial conditions.

As a final test to make sure that interactions between the subclusters (e.g. the close encounter between sc1 and sc3) are not important to the expansion we see, we run each subcluster in isolation. There are no notable differences to the main run. We plot the half-mass radii of the isolated subclusters in Figure \ref{RescaledClusters} below; they are consistent with the spread of the randomized runs.

\subsubsection{A note on the initial cluster radii}
Pure \nbody simulations are scale free. If one scales up
the initial radius of a cluster by a factor $f$ then the evolution
timescale is multiplied by a factor $f^{3/2}$. The present
simulations are, however, not entirely scale free because of the
inclusion of mass loss associated with stellar winds and supernovae:
the former is a mild influence on the cluster evolution while the
latter typically occur at $> 5$ Myr for clusters numbering
hundreds of stars. Therefore up to this point the simulations are roughly
scale free, and we can explore the effect of changing
the initial cluster radii by applying the above scalings.

\begin{figure}
 \includegraphics[width=84mm]{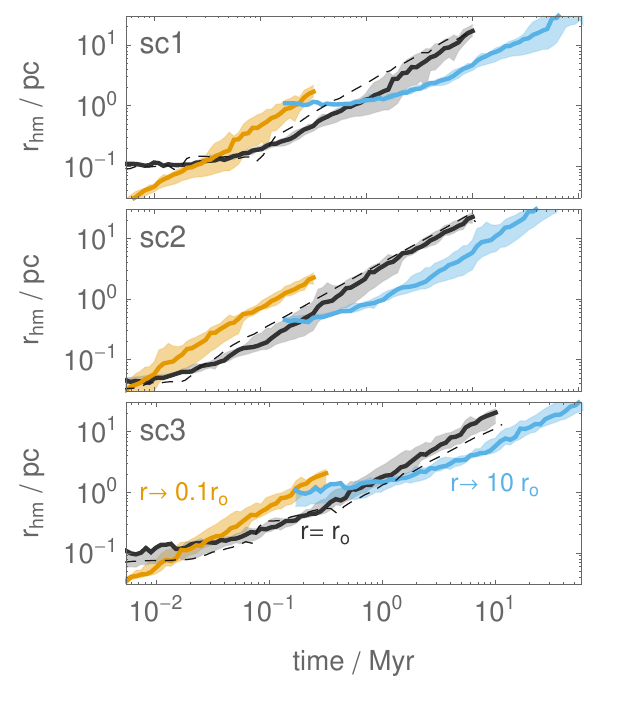}
 \caption{The half-mass radii of the initially bound stars, represented as the median of the randomised initial conditions. The simulations have been rescaled to initial half-mass radii an order of magnitude smaller (yellow lines) and larger (blue lines) than the base simulations (black lines). The filled bands show the maximum and minimum of the random simulations. The dashed lines show the result of running the subclusters in isolation.}
  \label{RescaledClusters}
\end{figure}

Figure \ref{RescaledClusters} shows the evolution of the half mass radius of
initially bound stars from the randomised simulations shown in
the lower panels of Figure \ref{RandomComparison}, and their scaled equivalents in
the case of clusters with initial radii that are a factor of
ten less than or greater than those modeled here. This range
in initial conditions corresponds to
six orders of magnitude in initial density. Nevertheless, by a
couple of Myr the evolution of the the rescaled clusters have all begun to transition to a common track
representing a self-similar expansion (i.e. the cluster has
expanded to the point that its age is always a fixed multiple
of its relaxation time). For these clusters containing
several hundreds of stars this transition has begun
by $\sim 2$ Myr for any cluster with an initial half mass
radius of a parsec or less.

Despite the orders of magnitude variation in initial size, the half-mass radii are within a factor of a few of each other by Myr timescales. Since the evolution of clusters of different
initial radii has largely converged by the time the first supernova
explodes, such clusters would be similarly affected by supernova
feedback. We thus conclude that our results are remarkably
insensitive to initial cluster radii

\section{Conclusions}
We have taken the end result of hydrodynamic simulation of star cluster formation and evolved its sink particles forward for 10 Myr after removing the gas entirely. Gas expulsion does not directly affect the subclusters in this simulation; they are virialised, stellar dominated and gas poor. Despite this the cluster expands from an overall half-mass radius of $\sim 1.6$ pc to $\sim 24$ pc, with a commensurate drop in the stellar surface density. This is due to two effects: the unbinding of the individual subclusters from each other as a result of the gas being removed, and -- dominantly -- the expansion of the individual subclusters driven by stellar dynamics and stellar evolution. While we focused here on the most massive and populous subclusters in the simulation, the same effects are taking place in the many other bound objects in the star forming region (Kruijssen et al. 2012 identify 20 subclusters in the hydro simulation). The less-populous groups will tend to evolve even more rapidly than those we analyse, and be even more susceptible to the chaotic, ejection-dominated behaviour seen in sc2.

The overall effect of the stellar dynamics is to make the subclusters expand to a similar degree as if they had been strongly affected by gas expulsion. A large fraction of the subcluster mass (something like a half or more, depending on the details of its evolution) escapes and streams to large radii, and the remaining bound mass has expanded to half-mass radii more characteristic of young clusters, of order 1 pc. We stress that these results only apply to clusters with small relaxation times, which typically means low-{\it N} clusters. The timescales for dynamical evolution are then short enough that there is essentially no equilibrium structure over the few Myr usually considered when thinking about the transition from an embedded to a gas-free cluster; rather there is a quasi-steady state of expansion and evaporation, and the density of the stars is a function of the cluster age rather then their initial conditions.

The evolution of the subclusters is consistent with the classical theory of an isolated cluster expanding due to energy generation at its core, in this case mostly from binary heating, punctuated by chaotic small-{\it N} dynamics that can free the remaining bound stars as the massive stars eject themselves.
The rapid dynamical evolution inherent in these low-mass, compact (thus high density) clusters implies that at the low-{\it N} end
of the cluster spectrum (i.e. the objects containing hundreds of stars
that dominate censuses of local star forming regions) the history of
cluster dispersion may be -- in contrast to what is generally assumed --
largely decoupled from the issue of the star formation efficiency. 

Moreover the stellar dynamical processes modeled here are, in contrast
to issues of gas expulsion, independent of the metallicity and
epoch of cluster formation. It is only in the case of significantly
more populous clusters with characteristically long relaxation times that gas expulsion and stellar dynamical
dispersal make significantly different predictions for cluster lifetimes.
In the case of these rarer high mass clusters, the need for cluster
infant mortality is largely addressed using extragalactic systems
\citep[see, for example,][]{bastian05b, maschberger11,chandar10b,bastian12}
In these cases, however, the interpretation is affected by
factors such as the assumed star formation history
and the effects of age dependent sensitivity limits so that the
observational
requirements on infant mortality in the high mass range remain
controversial.

\section*{Acknowledgments}
Our thanks to Mark Gieles, Diederik Kruijssen, and Thomas Maschberger for comments. Simon Goodwin's quick and constructive referee's report helped strengthen the paper.

%\bibliographystyle{mn2e}	
%\bibliography {/Users/nickolasmoeckel/Documents/astronomy/astro-library/allrefs}

\bsp

\label{lastpage}

\end{document}